\documentclass[useAMS,usenatbib]{mn2e}
\usepackage{graphicx}

\title[A variable Quasi-Periodic Oscillation in M82 X-1]
{A variable Quasi-Periodic Ocillation in M82 X-1.\\ Timing and
spectral analysis of {\it XMM-Newton} and {\it RossiXTE}
observations\thanks{Based on observations obtained with {\it XMM-Newton}, 
an ESA science mission with instruments and contributions directly funded 
by ESA Member States and NASA, and with {\it RossiXTE}.}}

\author[P. Mucciarelli et al.]{P. Mucciarelli$^{1,2}
$\thanks{E-mail:mucciarelli@pd.astro.it}, P. Casella$^{3}$, T. Belloni
$^{3}$, L. Zampieri$^{1}$, P. Ranalli$^{4}$
\\
$^{1}$INAF-Osservatorio Astronomico di Padova, Vicolo dell'Osservatorio 5,
I-35122, Padova -Italy\\
$^{2}$Dipartimento di Astronomia, Universit\`a di Padova, Vicolo dell'Osservatorio 2,
I-35122, Padova -Italy\\
$^{3}$INAF-Osservatorio Astronomico di Brera, Via E. Bianchi 46, I-23807 Merate
(LC), Italy\\
$^{4}$INAF-Osservatorio Astronomico di Bologna, Via Ranzani 1, I-40127 Bologna,
Italy}

\begin{document}

\date{
Accepted ... Received ...; in original form ...}

\pagerange{\pageref{firstpage}--\pageref{lastpage}} \pubyear{2002}

\maketitle

\label{firstpage}

\begin{abstract}
We report results from a spectral and timing analysis of M82 X-1,
one of the brightest known ultraluminous X-ray sources. Data from a
new 105 ks {\it XMM-Newton} observation of M82 X-1, performed in April
2004, and of archival {\it RossiXTE} observations are presented. A
very soft thermal component is present in the {\it XMM}
spectrum. Although it is not possible to rule out a residual contamination
from the host galaxy, modelling it with a standard
accretion disk would imply a black hole mass of $\approx 10^3
M_{\odot}$.  An emission line was also detected at an energy typical
for fluorescent Fe emission.
The power density spectrum of the {\it XMM} observation
shows a variable QPO at frequency of 113 mHz with properties similar
to that discovered by \cite{sm03}. The QPO was also found in 7
archival {\it RXTE} observations, that include those analyzed by 
\cite{sm03} and \cite{ft04}. 
A comparison of the properties of this QPO with those of the various
types of QPOs observed in Galactic black hole candidates strongly suggests
an association with the type-C, low frequency QPOs. Scaling the frequency
inversely to the black hole mass, the observed 
QPO frequency range (from 50 to 166 mHz) would yield a black hole mass
anywhere in the interval few tens to 1000 $M_\odot$.
\end{abstract}


\section{Introduction}

First revealed by the Einstein Observatory (see e.g. \citealt{fab89}),
point-like, off-nuclear X-ray sources with luminosities significantly
in excess of the Eddington limit for one solar mass have been
discovered in a large number of nearby galaxies
(e.g. \citealt{cp02,cm04,sw04,lb05}). The nature of most of these ultraluminous
X-ray sources (ULXs) remains unclear. In several cases a variability
on timescales of months/years has been detected, a possible signature
of an accreting compact object. Were these sources X-ray binaries in
the host galaxy, and assuming Eddington-limited accretion, masses in
the range 100-1000 $M_\odot$ are inferred from the observed flux. The
central object would be then an intermediate mass black hole
(IMBH). Mass estimates based on the Eddington limit are, however,
questionable: these sources need not to be spherically symmetric, nor
stationary. It has been proposed that many of the ULX properties can
be explained assuming that they do not emit isotropically \citep{ki01}
or are dominated by emission from a relativistic jet
(e.g. \citealt{ka03}). In this case, they may harbor stellar mass BHs
and may be similar to Galactic black-hole binaries.

The available {\it XMM-Newton} and {\it Chandra} observations show
that, for sources with sufficiently good statistics, the best fit of
the X-ray spectrum is often obtained with a two-component model: an
absorbed multicolor disk (MCD)+power-law. Some ULXs have MCD
temperatures $k T_{MCD} \sim 200-300$ eV, much lower than in Galactic BH
binaries. Since $T_{MCD} \propto M_{BH}^{-1/4}$, the low temperature
of the soft component has been taken as further evidence for the
presence of an IMBH of $\sim$ 100--1000 $M_\odot$ in some ULXs
\citep{mi03,mi04,cr04,zamp04,kong05}. A few sources have a temperature 
above 1 keV, higher than that of Galactic BH candidates, but still in the
range of Galactic microquasars.

Optical counterparts of ULXs have been studied by a number of authors
(see e.g. \citealt{zamp04} and \citealt{liu05}, and references
therein). Several ULXs are associated with optical emission nebulae of
a few hundred parsecs in diameter, whose properties are consistent
with those of X-ray irradiated nebulae \citep{pm02}.

Another approach to study the nature of ULXs is through time
variability.  The analysis of the aperiodic variability in the X-ray
flux of X-ray binaries is a powerful tool to study the properties of 
the inner regions of the accretion disk around compact objects 
(for a review see \citealt{vdk05}).  In particular, Quasi-Periodic
Oscillations (QPOs) provide well-defined frequencies, which can be
linked to specific time scales in the disk. QPOs can be broadly
divided into three classes: (a) QPOs at very low frequencies ($<$0.02 
Hz), probably associated to oscillations and instabilities in the
accretion disk (see \citealt{mo97,b97,b00}); (b) Low-Frequency (LF)
QPOs, with typical frequencies between 0.1 and 10 Hz, probably
connected to similar oscillations in neutron star systems (see
e.g. \citealt{b02,r02a,vdk05,cas05}), over whose origin there is no
consensus; in Black Hole Candidates (BHCs) 3 main different types of 
LF QPOs have been identified (\citealt[][and references therein]{cas05}); 
(c) ``hecto-Hertz'' QPOs, with a typical
frequency of 100-300 Hz, in two cases observed to appear in pairs
\citep{st01a,st01b}. It is currently unclear whether these QPOs show a
constant frequency for each source (see \citealt{h01,r02b}), and
whether they do appear in pairs obeying particular frequency ratios
(see \citealt{r02b}). However, since they identify the highest
frequencies observed in these systems, they are the best candidates
for association with, e.g., the keplerian frequency at the innermost
stable orbit, or relativistic precession frequencies.

Whatever their physical nature, as they originate in the inner regions
of accretion disks around black holes, these features are expected to
be produced also in ULXs. However, if ULXs contain IMBHs of 100-1000
$M_\odot$, the frequencies involved are much smaller. The first and,
to date, only ULX where a QPO has been discovered is M82 X-1
\citep{sm03}. The QPO has a frequency of 54.4 mHz and a FWHM of 11.4
mHz, leading to a quality value $Q=\nu / \Delta\nu \sim$ 5.  The
total fractional rms of the QPO in the 2-10 keV band is
8.4\%. Recently, \cite{ft04} reported the identification of another QPO
at 106 mHz in the power spectrum of M82 X-1 from {\it RossiXTE} data,
arguing that it may be a harmonic of the QPO at 54 mHz.

Here we report the results from a joint spectral and timing analysis
of a 105 ks {\it XMM-Newton} observation of M82 X-1 performed in April
2004 and of archival {\it RossiXTE} observations of the same
field. The new {\it XMM} data are part of a pointing to M82,
originally requested to study the enrichment of the galactic
interstellar medium (Ranalli et al., in preparation). The plan of 
the paper is the following. In \S \ref{spec} we present the spectral 
analysis of the new {\it XMM} observation and a re-analysis of the data by
\cite{sm03}. \S \ref{lc} reports the long term light curve of M82 X-1.
In \S \ref{timing} the timing analysis of the new {\it XMM}
observation and other archival {\it RossiXTE} observations is
presented. Finally, in \S \ref{disc} we discuss our results.

\begin{figure*}
 \begin{center} \includegraphics[]{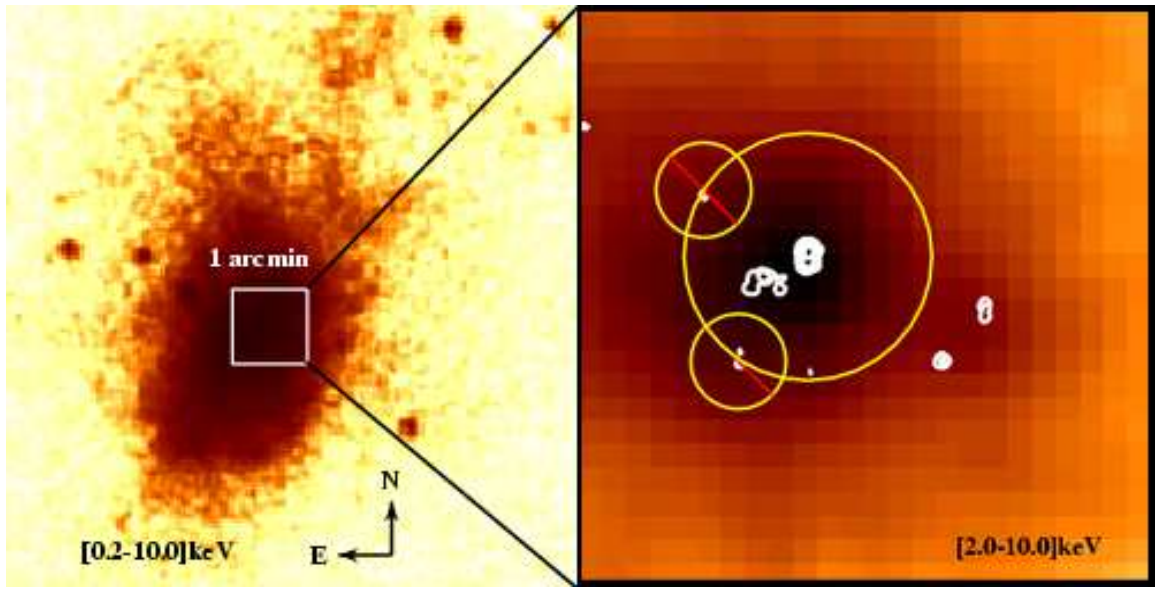} \end{center}
 \caption{{\em XMM} EPIC MOS1 exposure of M82, taken on April 21,
 2004. {\it Left}: full image of the galaxy in the [0.2-10.0] keV
 band. {\it Right}: 1$' \times$1$'$ region centered on M82 X-1 in the
 energy range [2.0-10.0] keV. Contour plots show the position of all
 the sources resolved in the {\it Chandra} observation of January 2000
 \citep{mats01}. The big circle (13$''$) shows the region used
 for source count extraction, while the two smaller (5$''$) circular
 regions are used to subtract the contribution of sources 2 and 3 of
 Matsumoto et al. (2001).} \label{fig:m82field}
\end{figure*}

\section[]{Spectral Analysis}
\label{spec}

\subsection{{\it XMM-Newton} observation of April 2004}

{\it XMM-Newton} observed the galaxy M82 on April 21, 2004
(Observation ID 0206080101, PI: P. Ranalli) for a total exposure time
of 105 ks. The three {\it XMM} EPIC cameras (pn, MOS1 and MOS2)
operated in Full Frame mode with the medium filter. Photon lists, data
screening, region selection and events extraction were performed with
standard software ({\sc XMM-SAS} v 5.4.1 and 6.0.0). Events lists were
directly extracted from the observation data files using the tasks
{\tt epproc} and {\tt emproc} for the EPIC pn and MOS data,
respectively.  The last part of the exposure was affected by high
background radiation. Standard high flares filtering (total off-source
count rate above 10 keV less than 1 count s$^{-1}$ for pn and 0.35
count s$^{-1}$ for MOS) leaves 76/72 ks of good time intervals for
the pn/MOS cameras. Slightly different good time intervals were used
for the spectral and timing analysis, as specified below.

Source counts of M82 X-1 were extracted from a circular region of
radius 13$''$, centered on the coordinates RA= 09:55:50.2,
DEC=+69:40:47. A previous observation with {\it Chandra} showed
several sources in or nearby the position of M82 X-1
\citep{mats01}. To avoid as much as possible contaminations we
subtracted the contribution of some of them by excluding circular
regions of 5$''$ radius centered on the {\it Chandra} positions
(see Figure \ref{fig:m82field}). The nearest sources could not be
eliminated in this way. Another strong source of background
contamination, especially at low energies, is the host galaxy
itself. \cite{sm03} did not attempt to subtract it and limited their
analysis to energies $\ge 2$ keV. Here we try to perform spectral fits
of the {\it XMM} data in the [0.8-10] keV range and therefore perform
subtraction of the host galaxy diffuse emission. The background was
extracted from an annulus of inner and outer radii of 18 and 35$''$,
respectively, eliminating the contribution of sources 1, 8 and 9 of
\cite{mats01} in a way similar to that described above. The background
subtracted count rate is $\sim$1.0 count s$^{-1}$ for pn and
$\sim$0.35 count s$^{-1}$ for each MOS.

\begin{figure}
 \begin{center}
	\includegraphics[angle=0,width=8.5cm]{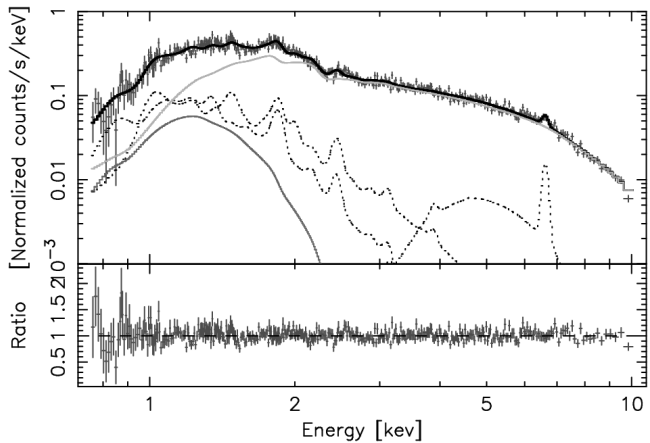}
 \end{center} 
\caption{{\it XMM} EPIC pn spectrum of M82 X-1 in April 2004 (Obs. ID 0206080101) with the best fitting 
model ({\it heavy solid line}) and components: PL ({\it gray solid line}) +
MEKALs ({\it thick dotted lines}) + MCD ({\it thick solid line}).}
\label{fig:m82fitnew}
\end{figure}

\begin{figure}
 \begin{center}
	\includegraphics[angle=0,width=8.5cm]{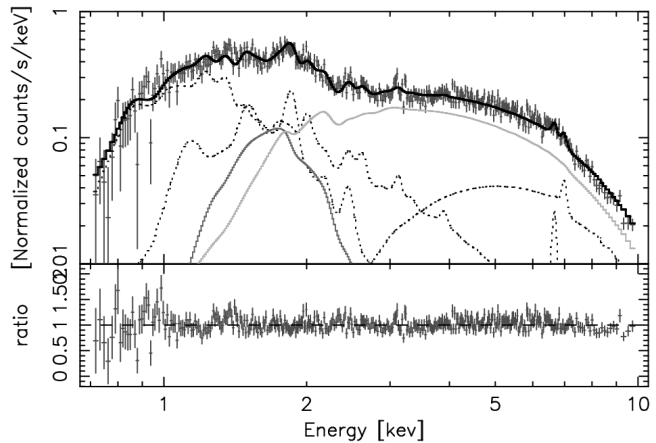}
 \end{center} 
\caption{Same as Figure \ref{fig:m82fitnew} for the May 2001 observation (Obs. ID 0112290201).}
\label{fig:m82fitold}
\end{figure}

Spectra were grouped to require at least 150/80 counts per bin for
pn/MOS and analyzed with XSPEC v.11.2.0. Standard interstellar
absorption ({\sc wabs}) was taken into account and an overall
normalization constant was used to minimize residual detector
calibration uncertainties. Results of the spectral fits
with various models are
reported in Table \ref{tab:m82reg}. The spectrum is rather composite,
so that fits with single component models do not give satisfactory results.
In particular, the residual contribution from the host galaxy
at soft energies ($<$ 2 keV) needs to be adequately modelled.
In a different context, \cite{stev03} tried to fit the nuclear emission
of M82 with two MEKAL (optically thin, thermally emitting plasma)
components plus a power-law (PL). Following \cite{stev03},
we tried to fit the residual galactic contamination with a dual MEKAL model
with different absorbing columns. Using this model for the diffuse emission
plus an absorbed PL for the ULX gives a reasonable fit ($\chi^{2}=1.31$ for 846 d.o.f.). 
This can be further improved adding a third MEKAL component, needed mainly
to fit a residual excess in emission around 6.6 keV ($\chi^{2}=1.24$ for 843 d.o.f.).
This line, attributed to fluorescent K$\alpha$ Fe emission, can also be fitted
adding a gaussian to the PL component.

\begin{table*}
\begin{tabular}{l|ccccc}
\hline
Model&Column density
&Temperature&Photon index& Line param.&\\
&$\frac{N_{H}}{10^{22} cm^{-2}}$&$\frac{kT}{keV}$&$\Gamma$&$\frac{energy, width}{keV}$&Red. $\chi^{2}
(dof)$\\
\hline        
\multicolumn{6}{l}{Obs.ID 0206080101}\\
\hline                                                                              
      
WABS*MEK +                & 1.66$\pm$0.08&0.51$\pm$0.03 &				  &&	     \\
WABS*MEK +                & 1.31$\pm$0.18&0.99$\pm$0.07 &				  &&	     \\
WABS*PL                   & 1.38$\pm$0.07&		   &1.54$\pm$0.02	 &&1.31(846)\\
\hline                          						   
      
WABS*MEK +                & 1.67$\pm$0.08&0.51$\pm$0.04 &				 &&	    \\
WABS*MEK +                & 1.31$\pm$0.19&0.96$\pm$0.07 &				 &&	    \\
WABS*(PL+GAUSS)           & 1.38$\pm$0.07&&1.56$\pm$0.02 &E=6.64$\pm$0.03&1.27(843)\\
                        &       	        &&&$\sigma=$0.09$\pm$0.01&\\
\hline                          	        				  
      
WABS*MEK +                & 1.67$\pm$0.10&0.49$\pm$0.06 &				 &&	    \\
WABS*MEK +                & 1.44$\pm$0.20&0.94$\pm$0.08 &				 &&	    \\
WABS*MEK +                & 8.86$\pm$4.34&7.59$\pm$1.22 &				  &&	     \\
WABS*PL                   & 1.31$\pm$0.11&			    
&1.61$\pm$0.05        &&1.24(843)\\
\hline                                                                              
      
WABS*MEK +                & 1.97$\pm$0.34&0.43$\pm$0.06 &				 &&	    \\
WABS*MEK +                & 1.47$\pm$0.16&0.90$\pm$0.03 &				 &&	    \\
WABS*MEK +                & 9.96$\pm$4.64&8.24$\pm$1.21 &				  &&	     \\
WABS*(PL+MEK)             & 1.39$\pm$0.10&0.14$\pm$0.03
&1.65$\pm$0.05        &&1.24(841)\\
\hline        
WABS*MEK +                & 1.60$\pm$0.12&0.48$\pm$0.08 &				  &&	     \\
WABS*MEK +                & 1.68$\pm$0.18&0.92$\pm$0.07 &				  &&	     \\
WABS*MEK +                &13.72$\pm$10.04&6.35$\pm$1.44 &				    &&         \\
WABS*(PL+MCD)             & 1.76$\pm$0.24&0.18$\pm$0.01
&1.66$\pm$0.06        &&1.22(841)\\
\hline                                                                              
      
\multicolumn{6}{l}{Obs.ID 0112290201}\\
\hline                                                                              
      
\hline        
WABS*MEK +                & 1.50$\pm$0.04&0.56$\pm$0.01 &				  &&	     \\
WABS*MEK +                & 1.82$\pm$0.14&1.67$\pm$0.11 &				  &&	     \\
WABS*(PL+MCD)             & 5.99$\pm$0.31&0.19$\pm$0.01
&1.53$\pm$0.04        &&1.17(884)\\
\hline                                                                              
      
\hline 
\end{tabular}        
\caption[]{\footnotesize Parameters from the spectral fits of M82 X-1 in the
[0.8-10.] keV band.
Results are presented for our new {\it XMM-Newton} observation
(Obs. ID 0206080101) and for the archival observation of \cite{sm03}
(Obs. ID 0112290201). The data were extracted from a region of 13$''$ radius for both
observations (see text for details).}
\label{tab:m82reg}
\end{table*}
Adding an additional soft component to the PL results in a further improvement 
of the fit. 
The best fit of the joint pn and MOS spectra of M82 X-1 is obtained adding
to the composite 3-MEKAL model of the diffuse galactic emission and the PL
a multicolor disk (MCD) component with a very low temperature
($kT=180$ eV; see again Table \ref{tab:m82reg}). Adding the MCD component to the
3-MEKAL plus PL model leads to an improvement in $\chi^{2}$ of $\sim 20$. 
A plot of the [0.8-10.0] keV
EPIC pn spectrum with the best fitting spectral model is shown in Figure \ref{fig:m82fitnew}.

To check the accuracy of the power-law photon index, we tried to fit
only the high energy part of the spectrum of M82 X-1 in the energy
interval [3.3-10.0] keV
(cfr. \citealt{ft04}). This should clearly minimize the contamination
from the galactic diffuse soft emission. In this energy range
the fits were performed with a power-law plus one or two MEKAL
components. Without fixing the value of the column density, unacceptable
fits or unreasonably high/low values of some parameters are obtained.
Thus, we kept $N_H$ fixed at the values derived from the best fits in
Table \ref{tab:m82reg}. The resulting photon
index is in the range $\Gamma=1.73-1.80$, in agreement within $2\sigma$
with that derived from the full spectral fit, $\Gamma= 1.66$ (see Table
\ref{tab:m82reg}).

\subsection{{\it XMM-Newton} observation of May 2001}

The {\it XMM} observation taken on May 6, 2001 (Observation ID
0112290201), analyzed by \cite{sm03}, lasted 31 ks. We re-analyzed it
following the procedure outlined in the previous subsection. Standard
flares filtering leaves 22 and 29 ks of good time intervals for the pn
and MOS cameras, respectively. Source counts were extracted from a
13$''$ region centered on the same source position. The net count rate
is $\sim$1.6 count s$^{-1}$ for pn and $\sim$0.56 count s$^{-1}$
for each MOS. The best fit of this observation is obtained with a dual
MEKAL model plus a PL+MCD with different absorbing columns. 
Adding a third MEKAL component leads to an improvement of the fit, but the 
temperature converges to unphysically large values ($\sim 100$ keV).
The parameters of the best fit are reported in Table \ref{tab:m82reg}.
The temperature of the MCD component and the PL photon index are
consistent, within $2 \sigma$, with those of the April 2004 observation. 
However, the absorbing column of these components is approximately 3
times larger than that measured in the 2004 observation. It should be 
noted that the relative normalization and temperatures of the MEKAL 
components vary slightly between the 2001 and 2004 observations. Clearly, 
this is not a consequence of a variation in the diffuse galactic emission, 
but reflects simply the fact that the model adopted for it is approximate.

Also in this case, we checked the accuracy of the PL photon index
fitting the spectra in the restricted energy interval [3.3-10.0] keV.
The fits were performed with a PL, with or without
a MEKAL component. As for the 2004 observation, without freezing $N_H$
unacceptable fits or unreasonably high/low values
of some parameters are obtained. Fixing the column density at the value
reported in Table \ref{tab:m82reg}, the resulting photon
index is in the range $\Gamma=1.66-1.68$, consistent within $3 \sigma$
with that of the full spectral fit, $\Gamma=1.53$.

Finally, we note that, although \cite{sm03} reported evidence for a Fe line at 6.55 keV
from an analysis of the spectral residuals, we are unable to confirm
its detection (the improvement obtained adding a gaussian component 
to the fit is negligible).

\begin{table*}
\begin{tabular}{l|cccc}
\hline
Observatory	& Date	& Count Rate	& Absorbed Flux$^a$			&
Unabsorbed flux$^a$ \\
      		&	&(count s$^{-1}$)&(erg cm$^{-2}$ s$^{-1}$)	&(erg
		cm$^{-2}$ s$^{-1}$)\\
\hline
{\it Chandra} HRC-I	&1999-10-28	&0.07$^b$		&$7.8\times 10^{-12}$/$5.1\times 10^{-12}$ $^c$&$3.9\times
10^{-11}$/$1.4\times 10^{-11}$ $^c$\\
{\it Chandra} HRC-I	&2000-01-20	&0.52$^b$		&$5.8\times 10^{-11}$/$3.8\times 10^{-11}$ $^c$&$2.9\times
10^{-10}$/$1.1\times 10^{-10}$ $^c$\\
{\it XMM} EPIC		&2001-05-06	&1.6/0.56$^{d}$ &$2.4\times
10^{-11}$&$1.2\times 10^{-10}$\\
{\it XMM} EPIC		&2004-04-21	&1.0/0.35$^{d}$ &$1.2\times 10^{-11}$&$3.4\times 10^{-11}$\\
\hline
\multicolumn{5}{l}{$^{a}$ For the {\it XMM} EPIC observations,
average of the pn and MOS fluxes as calculated by XSPEC}\\
\multicolumn{5}{l}{$^{b}$ From \cite{mats01}}\\
\multicolumn{5}{l}{$^{c}$ See text for details about the adopted spectral model}\\
\multicolumn{5}{l}{$^{d}$ pn/MOS count rates}\\
\hline
\end{tabular}	
\caption[]{\footnotesize The 0.2-10 keV fluxes of M82 X-1 from the {\it Chandra} and {\it XMM}
observations.
}
\label{tab:longlc}
\end{table*}

\section{X-ray light curve}
\label{lc}

Table \ref{tab:longlc} shows the long time scale flux variability of
M82 X-1 from {\it XMM} and {\it Chandra} data. Measurements from {\it
RXTE} are not reported because it does not have imaging capabilities
and contamination from the host galaxy and nearby point sources may
significantly affect the count rate. For the {\it XMM} observations
the fluxes are those of the best fit models of Table \ref{tab:m82reg},
avaraged over the three EPIC instruments. For the {\it Chandra} HRC--I
observations, the fluxes are calculated using the web interface to
PIMMS (v. 3.6c). We adopt a simplified MEKAL and MCD+PL model with
two different absorbing columns and approximate the MEKAL/MCD
component with a Bremsstrahlung/Blackbody at the same temperature. The
{\it Chandra} fluxes are probably underestimated because the shape of
the MCD and blackbody spectra start to differ significantly below 0.5
keV.

The highest recorded flux emitted by M82 X-1 ($F \sim 10^{-10}$ erg
cm$^{-2}$ s$^{-1}$) corresponds to a luminosity $L\sim 2\times
10^{41}$ erg s$^{-1}$ ($D=3.9$ Mpc; \citealt{sama99}). Making the
usual assumption that, at maximum, the source emits at the Eddington
limit, we can derive a rough estimate of the BH mass, $M_{BH} \sim
1500 M_{\odot}$.

\section{Timing analysis}
\label{timing}

\subsection{{\it XMM-Newton} data}

\begin{figure}
 \begin{center}
	\includegraphics[width=8.5cm]{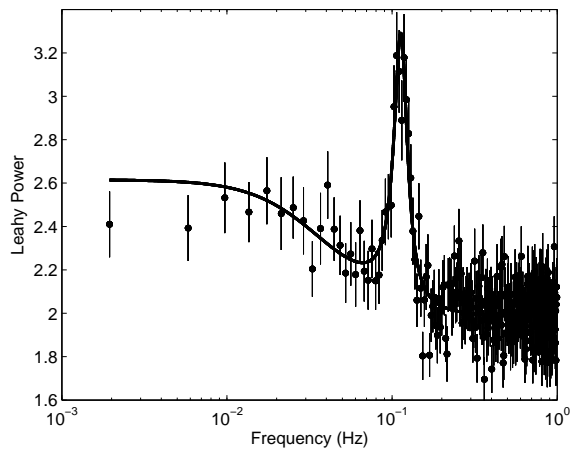}
 \end{center} 
\caption{Total power spectrum from the 2004 {\it XMM} observation (limited to the range 0.001-1 Hz).
	The line represents the best fit model (see text).}
\label{fig:pds}
\end{figure}

For the timing analysis of the 2004 data we avoided the interval with
high background radiation and limited the extraction to the longest
(nearly) uninterrupted segment of data (66 ks) free from solar flares
with count rate higher than 30 count s$^{-1}$.
To minimize galactic contamination, source counts were extracted from
a circular region of 8$''$ radius and at energies $>$ 2 keV. We
produced a light curve from pn+MOS data with a time binning of 0.5
s. A few gaps of typical duration of $\sim$100 s were present in the
light curve and were filled with a Poissonian realization around the
mean value of counts before and after the gap. We produced a power
spectrum (normalized after \citealt{le83}) from the resulting light
curve and rebinned it by a factor of 256 reaching a frequency
resolution of 3.9 mHz (see Figure \ref{fig:pds}).  A rather strong QPO
peak is evident in the figure. We fitted the power spectrum with a
model consisting of a constant (for the Poissonian level) plus two
Lorentzian components (see \citealt{b02}): one zero-centered for the
broad band-limited noise and one for the QPO peak. The characteristic
frequency for the band-limited noise component (see \citealt{b02} for
a definition) is 39.4$\pm$8.6 mHz and its integrated fractional rms is
$\sim$22\% (after subtracting the contribution of the host galaxy). 
The parameters of the QPO can be seen in
Table \ref{tab:qpo}. The quality value $Q$, defined as the ratio of the
centroid frequency over the FWHM of the QPO, is 4.3$\pm$0.5.  We
repeated the analysis in two separate energy bands, 2-4 keV and 4-10
keV.  The fractional rms of the QPO in these bands resulted to be
13.8\% and 23.9\% respectively.

\begin{figure}
 \begin{center}
	\includegraphics[width=6.5cm]{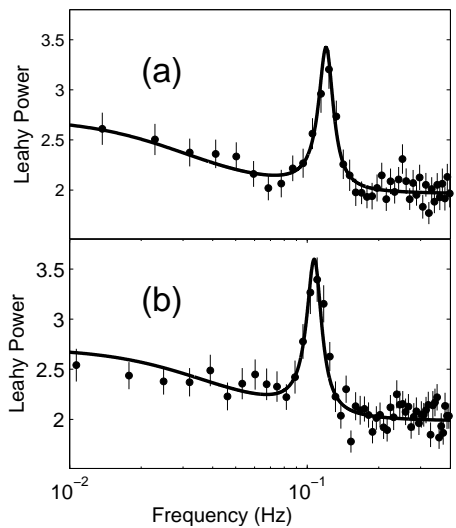}
 \end{center} 
\caption{(a) Power spectrum from the first half of the {\it XMM} data. (b) Power
	spectrum from the second half. The lines show the best fit models
	described in the text.
}
\label{fig:pds_comparison}
\end{figure}

In order to investigate the possible
variability of the QPO during the observation, we produced a spectrogram,
by aligning power spectra obtained from consecutive stretches of data 
2048 seconds long. The spectrogram is shown in Figure \ref{fig:dyna}.
For display purposes, in the spectrogram each pixel is
smoothed as the bilinear interpolation of the values at its vertices.
A trend towards lower QPO frequencies is apparent, correlated
with the source count rate (top panel). In order to quantify the decrease
in centroid frequency,
we divided the 66 ks interval in two segments of 33 ks each and repeated the
power spectral analysis described above. The two resulting power spectra
are shown in Figure \ref{fig:pds_comparison}. A fit with the same model used
for the total power spectrum confirms that the centroid frequency of the QPO
decreased by 10.8$\pm$4.0\% (see Table \ref{tab:qpo}).

\begin{table*}
\begin{tabular}{l|ccc}
\hline
Parameter       & Total observation     & First half            & Second half \\
\hline
$\nu_0$ (mHz)   &     113$\pm$2         &    120$\pm$3          & 107$\pm$4 \\
$ FWHM (mHz)$   &      26$\pm$3         &     21$\pm$4          &  19$\pm$3 \\
Frac \% rms     &      18.3$\pm$1.0     &     17.5$\pm$1.1      &  17.3$\pm$1.1 \\
Signif. ($\sigma$)&     8.9             &      8.3              &   8.2     \\
\hline
\end{tabular}	
\caption[]{\footnotesize Parameters of the {\it XMM} QPO. Errors are at 1$\sigma$
level.
}
\label{tab:qpo}
\end{table*}

\begin{figure}
 \begin{center}
 \end{center} 
\caption{Spectrogram obtained from 2048 s {\it XMM} data stretches. Darker gray corresponds
to higher power. [{\bf This Figure is available as a separate jpg file}].
}
\label{fig:dyna}
\end{figure}

In Galactic BHCs the frequency of some QPOs is correlated with certain
spectral parameters, in particular with the power law spectral index
(see e.g. the QPO frequency-$\Gamma$ relation in GRS 1915+105;
\citealt{vig03}). We then analyzed the spectra of M82 X-1 in the two
33 ks intervals after filtering for solar flares, adopting the best
fitting model reported in Table \ref{tab:m82reg}. The spectral analysis does not show any
evidence of variability in the PL spectral index in the two intervals.

We also re-analyzed the {\it XMM} observation 0112290201 of
\cite{sm03}. The analysis of the power density spectrum confirms the 
presence of the QPO found by \cite{sm03}.

\subsection{{\it RXTE} data}

\begin{figure}
 \begin{center}
	\includegraphics[width=9.5cm]{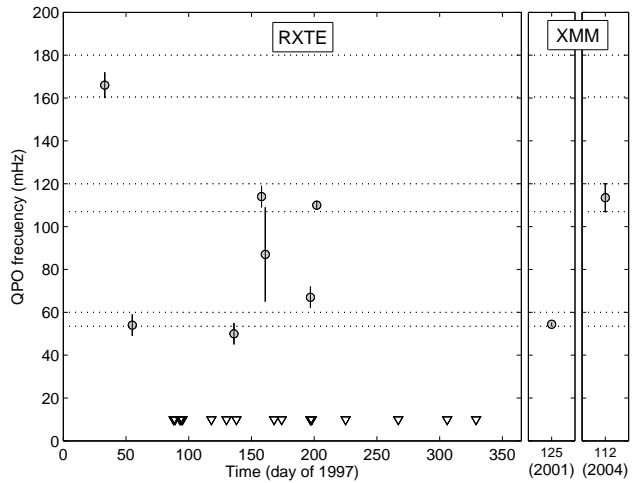}
 \end{center} 
\caption{Time history of the centroid frequencies detected from M82 X-1
	in the {\it XMM} and {\it RXTE} data.
	The triangles indicate the times of {\it RXTE} observations when no significant
	QPO was detected. The pairs of dotted lines indicate the range of
	frequencies detected by {\it XMM} in 2004 and the corresponding intervals
	at half and 1.5 times the frequency.
}
\label{fig:rxte}
\end{figure}

In order to investigate the variability of the QPO frequency on longer time
scales, we extracted from the {\it RXTE} public archive all 30 public observations
of M82, spanning over the year 1997. For each observation, we accumulated 
PCA light curves in the channel range 0-35, corresponding to 2-13 keV,
with a 0.5 s bin size and produced power spectra in the same way as for the
{\it XMM} data. We detected a significant QPO in seven observations, including
the three reported by \cite{sm03} and \cite{ft04}. These detections are summarized
in Table \ref{tab:rxte} and a timing history of their centroid frequencies
is shown in Figure \ref{fig:rxte}, where also the {\it XMM} detections are
indicated. Although the frequencies are variable,
they are roughly consistent with three groups in harmonic 1:2:3 ratio, as
recently suggested by \cite{ft04}.
In order to calculate the significance of such an harmonic relation we did a
numerical simulation and found the nine QPO frequencies (the seven from RXTE
data plus the two from XMM) to be consistent at 2.8 $\sigma$ with being
harmonics of a fundamental frequency of 54.9 Hz.
However, this is not sufficient to completely rule out that such a distributions occurs 
by chance and then we did not explore further the consequences 
of this result. More detections are clearly needed in order to address 
this issue.

\begin{table*}
\begin{tabular}{l|cccc}
\hline
ObsID         & Date       & Exp. (s)& QPO $\nu$ (mHz)  & $n_\sigma$  \\
\hline
20303-02-01-00& 1997 Feb 02&   3709  &     166$\pm$6    &  6.8        \\
20303-02-02-00& 1997 Feb 24&   3616  &      54$\pm$5    &  5.3        \\
20303-02-03-00& 1997 May 16&   3312  &      50$\pm$5    &  6.0        \\
20303-04-04-00& 1997 Jun 07&   3872  &     114$\pm$5    &  5.8        \\
20303-04-05-00& 1997 Jun 10&   2848  &      87$\pm$22   &  4.4        \\
20303-08-07-00& 1997 Jul 16&   3127  &      67$\pm$5    &  6.5        \\
20303-02-04-00& 1997 Jul 21&   2896  &     110$\pm$2    &  4.3        \\

\hline
\end{tabular}	
\caption[]{\footnotesize Parameters of the {\it RXTE} QPO. Errors are at 1$\sigma$
level.
}
\label{tab:rxte}
\end{table*}

\section{Discussion}
\label{disc}

\subsection{Energy spectra}

M82 X-1 is one of the best studied ULXs. It suffers from huge
contamination in the soft X-rays from the diffuse emission of the host
galaxy, the starburst prototype M82 \citep{ori04}. Despite the care in 
subtracting the contribution of this emission, a residual contamination
is present in the spectra. 
We modelled this emission with a 3-MEKAL model with different absorbing columns.
The third MEKAL component at high temperature can account
for the He-like Fe excess emission in the spectrum, suggesting that it may 
originate from the diffuse emission in M82 and not from the ULX. The upper limit
to the flux of a line at $\sim 6.6$ keV in the 2001 XMM observation is consistent with
the line flux measured in the 2004 XMM observation. Since in 2001 the source count
rate was 50\% higher, it might have been difficult to detect a line with a flux
comparable to that of the 2004 observation. This result appears to be consistent with
the absence of line flux variability and then with a galactic origin. However,
it should be noted that the very high column density of the third MEKAL
component may represent a problem for this interpretation, although absorption 
columns as high as $10^{23}$ cm$^{-2}$ are reported in not-dissimilar galaxies 
\citep{piet01}.

The HRI instrument on board {\it Chandra} allowed \cite{mats01} to
resolve the central part of the galaxy, revealing the presence of a
number of point sources in addition to M82 X-1 (the {\it Chandra}
source CXOM82 J095550.2+694047, labelled source 7 by \citealt{mats01}).
A total of 9 sources are present in a field of $1'\times 1'$ centered
on the galaxy. Most of these sources are rather close to M82 X-1,
below the resolving power of the {\em XMM} EPIC cameras. For this
reason, as mentioned in \S \ref{spec}, we subtracted from the spectral
extraction region circular patches (5$''$) in coincidence with the position of
the {\it Chandra} sources.
Although we are aware that for {\em XMM} this is not as effective as for 
{\it Chandra} (because a region of 5$''$ corresponds to a {\em XMM} encircled 
energy fraction of about 30-50\%), this appears to be a reasonable compromise 
between minimizing the contamination from these sources and preserving
the counting statistics of M82 X-1. 
This procedure
could not be adopted to eliminate the contribution from the three
nearest sources to M82 X-1 (4, 5 and 6 in \citealt{mats01}). In
particular, source 5 is highly variable. In the {\it Chandra}
observation of October 1999 the count rate of this source was only
30\% of that of M82 X-1. Later, the former decreased by an
order of magnitude while the latter increased by a factor $\sim$
7. Therefore, while sources 4 and 6 are roughly constant and weak with
respect to M82 X-1, significant contamination from source 5 can not be
ruled out.

A number of ULXs show very soft spectral components well fitted by a
MCD model, representing emission from an accretion disk. Our best fit
includes a soft MCD component with temperature $kT_{MCD}$=0.18 keV.
A plot of the $\Delta \chi^2$ contours for $kT$ vs. normalisation of the 
MCD component shows that the temperature is well constrained while, within
3$\sigma$, the normalization varies between $\sim 500$ and $\sim 5000$ (XSPEC 
units; best fit value 1745). Although this is a large variation, the  
lower bound makes the detection of the MCD component quite robust.  
Adding this component results in a statistical improvement
of the fit. However, although a 3-MEKAL model with
different absorbing columns offers a fair representation of the starbust
diffuse emission, it is not possible to rule out completely a residual
contamination at very soft energies from the host galaxy.
Under the assumption that the softest emission is entirely due to an accretion
disk, the temperature of the MCD fit ($T_{MCD}$) can be used to
estimate the BH mass. 
We note that such an interpretation in terms of a standard accretion disk 
does not encounter the severe inconsistencies that are found when a similar 
hypothesis is applied to the soft excess in AGNs (see \citealt{gier04}).
Assuming that $T_{MCD}$
represents an estimate of the maximum temperature of a
standard accretion disk, it is $(3GM_{BH}{\dot M}/8\pi\sigma
r_{in}^3)^{1/4}=\alpha T_{MCD}$, with $\alpha\simeq 2$ (see \citealt{zamp04}). 
If the disk
terminates at the innermost stable circular orbit of a Schwarzschild
BH, the expression for the BH mass then becomes:
$M_{BH}/M_\odot=({\dot M}c^2/L_{Edd}) f^4 (\alpha T_{MCD}/1.5\times
10^7 \, {\rm K})^{-4}$, where $f\simeq 1.6-1.7$ is a color correction
factor \citep{shim95,zamp01}.  Assuming Eddington limited accretion,
the BH mass is $M_{BH} \approx 200 f^4 \, M_\odot$, roughly in
agreement with the value inferred from the X-ray flux.

\subsection{Timing}

M82 X-1 is at present the only ULX where a QPO has been
discovered. There are a total of 9 detections, two of which in the
{\it XMM} data and seven in the {\it RXTE} data, with frequencies in
the range $\sim$50-170 mHz. An important issue is of course the
possible identification of this QPO with one of the QPO types observed
in the X-ray light curves of stellar-mass BHCs. In the following we
will summarize the main properties of the QPO in M82 X-1, and discuss
its similarities and differences with the QPOs observed in BHCs.

\begin{itemize}

\item The lowest and highest observed frequencies are $50\pm5$
mHz and $166\pm6$ respectively (see Figure \ref{fig:rxte}).

\item The frequency distribution over this range is
suggestive of a harmonic 1:2:3 ratio between them. 

\item In the 2004 {\it XMM}
observation the frequency is observed to vary by a factor of $\sim$10\% in
less than one day.

\item The QPO peak has a quality value higher than 4 (up to $\sim$6 in one
case).

\item  It shows a high fractional rms (up to $\sim$18\%). 

\item The underlying band
limited noise is strong (fractional rms $\sim$22\%) and has a characteristic
frequency comparable to the QPO frequency. 

\item The integrated fractional rms of the QPO above 4 keV is higher than below
that energy.

\end{itemize}

Let us compare now all these properties with those of the various types of QPOs
observed in BHCs. 

\begin{itemize}

\item{\it Very-low frequency QPOs.}
An association of the QPO in M82 X-1 with the very low
frequency ($\nu\la 0.02$ Hz) ``heart-beat'' QPOs observed in GRS 1915+105 is
unlikely as their frequency is {\it lower}. This, assuming
an inverse scaling with the black-hole mass, would
imply a very low ($\sim$solar) mass black hole in M82 X-1, which is not in
agreement with the spectral evidences. 

\item{\it High-frequency QPOs.} The observed
short-time scale variability seems to exclude an association with the
high-frequency ``hecto-Hertz'' QPOs observed in BHCs, since the latter have
been detected at rather stable frequencies. 
However, it is worth noticing that, assuming a $10^3$ scaling factor between
the two phenomena, the $\sim 10$ mHz variation of the centroid frequency on a
time scale of $\sim$ 30 ks observed in M82 X-1 would correspond to a $\sim 10$ 
Hz variation on a time scale of $\sim$ 30 s in a 10 solar mass BHC.
Such a short time scale variability is impossible to detect in BHCs with 
the present instrumentation. On the other hand, the 
presence of a strong underlying band limited noise, with a characteristic 
frequency comparable to the QPO frequency, is clearly
at variance with the high-frequency "hecto-Hertz"QPOs observed in BHCs. 
Furthermore, the rms amplitude of the QPO itself in M82 X-1 is roughly an
order of magnitude bigger than that of the "hecto-Hertz" QPOs in BHCs, thus
making the association very unlikely. For the sake of
completeness we stress that the detection of the QPO at $\sim$ 166 mHz
reported in this paper lowers the upper limit for the mass of the black
hole in M82 X-1 (assuming that this frequency is associated with
the Keplerian frequency at the innermost circular orbit around a Schwarzschild
black hole) to $\sim 1.2 \times 10^4 M_{\odot}$.

\item{\it Low-frequency QPOs.} Three main types of low-frequency QPOs are observed in BHCs (see
\citealt{cas05}). In two of them, the type-A and the type-B QPOs, the
peak appears always at frequencies near 8 and 6 Hz respectively. Moreover,
they are both characterized by a weak (a few \%) underlying red
noise component. These properties make an association with the 
variable, strong
QPO observed in M82 X-1 unlikely. In the case of type-A QPOs, its low
coherence and amplitude make the association even less likely.

The properties of the QPO in M82 X-1 are on the contrary reminiscent
of those of the third type of BHCs low-frequency QPO, the type C, whose
characteristic frequencies vary in the range 0.1-15 Hz. The
similarities in fractional rms, variability, quality value, and 
underlying noise strongly
suggest an association between the two features. 
Furthermore, in the 2004 {\it XMM} observation there is evidence for a positive
correlation of the QPO frequency with the count-rate, and a similar
correlation is often observed in type-C QPOs. However, during the 2001 {\it XMM}
observation (when the QPO was detected at a lower frequency) the count rate
was higher than during the 2004 observation. Since the count-rate
vs. frequency correlation in BHCs is ``outburst dependent'' (which means
that during different outburst a source can show similar frequencies at
different count-rates) no conclusion can be derived from the observed
phenomenology in M82 X-1. No information on the count-rate variability could
be obtained from the {\it RXTE} observations, given the lack of imaging capabilities
of the satellite.


\end{itemize}



Assuming that the QPO detected in M82 X-1 is a type-C QPO, and scaling the
frequency inversely to the BH mass, the observed frequency range (from 50 to
166 mHz) would yield $M_{BH}$ anywhere in the range 10-1000
$M_\odot$. However, type-C QPOs are observed in BHCs throughout the whole
Hard-Intermediate State (see \citealt{hb05}), and their frequency is known to
decrease with the hardness of the energy spectrum. 
At the lowest observed frequencies,
the spectrum is hard and there is often no evidence for the presence of
a soft thermal component. As the contribution from a disk appears and
increases, the QPO frequency also increases. 
The {\it XMM} spectra of both observations
in which a QPO has been detected in M82 X-1 show possible evidence for a
disk contribution (see before). To the extent that the two phenomena can be
compared, the presence of a soft component would exclude 
that the type-C QPO in M82 X-1 is in the lowest frequency range, increasing the lower
limit for $M_{BH}$.

\section{Conclusions}

In this paper we reported a complete analysis of {\it XMM-Newton} and 
{\it RXTE} observations of the superluminal X-ray source M82 X-1. The
similarities in fractional rms, variability, quality value, and 
underlying noise strongly suggest an association between the QPO
in M82 X-1 and the low-frequency, type-C QPOs observed in BHCs.
The combined spectral
and timing analysis allows us for the first time to put strong constrains to
the mass of the central black hole in this source, yielding to a value
between a few tens to one thousand solar masses.

\section*{Acknowledgments}
We thank an anonymous referee for valuable comments.
This work has been partially supported by the Italian Ministry for
Education, University and Research (MIUR) under grants
PRIN-2002-027145 and PRIN-2003-027534\_004 and by INAF-PRIN grant.

\label{lastpage}

\end{document}